\documentclass[aps,prx,reprint,superscriptaddress]{revtex4-2}

\usepackage[T1]{fontenc}
\usepackage[utf8]{inputenc}
\usepackage{times,amsmath,amssymb,graphicx,enumitem}
\usepackage[dvipsnames]{xcolor}
\usepackage[colorlinks=true,allcolors=Blue]{hyperref}

\usepackage{xspace}

\usepackage{fancyhdr}
\setcitestyle{numbers, square}

\begin{document}

\newcommand{\ie}{{\it i.e.}\xspace}
\newcommand{\eg}{{\it e.g.}\xspace}
\newcommand{\etal}{{\it et al}.\xspace}

\newcommand{\Kxx}{$\kappa_{\rm xx}$}
\newcommand{\Kxy}{$\kappa_{\rm xy}$}
\newcommand{\TN}{$T_{\rm N}$}
\newcommand{\RuCl}{$\alpha$-RuCl$_3$}


\title{Evidence of a Phonon Hall Effect in the Kitaev Spin Liquid Candidate $\boldsymbol{\alpha}$-RuCl$_3$}
\author{\'E.~Lefran\c{c}ois}
\affiliation{Institut Quantique, D\'epartement de physique \& RQMP, Universit\'e de Sherbrooke, Sherbrooke, Qu\'ebec J1K 2R1, Canada}
\author{G.~Grissonnanche}
\affiliation{Institut Quantique, D\'epartement de physique \& RQMP, Universit\'e de Sherbrooke, Sherbrooke, Qu\'ebec J1K 2R1, Canada}
\author{J.~Baglo}
\affiliation{Institut Quantique, D\'epartement de physique \& RQMP, Universit\'e de Sherbrooke, Sherbrooke, Qu\'ebec J1K 2R1, Canada}
\author{P.~Lampen-Kelley}
\affiliation{Materials Science and Technology Division, Oak Ridge National Laboratory, Oak Ridge, Tennessee 37831, USA}
\affiliation{Department of Materials Science and Engineering, University of Tennessee, Knoxville, Tennessee 37996, USA}
\author{J.~Yan}
\affiliation{Materials Science and Technology Division, Oak Ridge National Laboratory, Oak Ridge, Tennessee 37831, USA}
\author{C.~Balz}\thanks{Present address: ISIS Neutron and Muon Source, STFC Rutherford Appleton Laboratory, Didcot OX11 0QX, United Kingdom.}
\affiliation{Neutron Scattering Division, Oak Ridge National Laboratory, Oak Ridge, Tennessee 37831, USA}
\author{D.~Mandrus}
\affiliation{Materials Science and Technology Division, Oak Ridge National Laboratory, Oak Ridge, Tennessee 37831, USA}
\affiliation{Department of Materials Science and Engineering, University of Tennessee, Knoxville, Tennessee 37996, USA}
\author{S.~E.~Nagler}
\affiliation{Neutron Scattering Division, Oak Ridge National Laboratory, Oak Ridge, Tennessee 37831, USA}
\author{S.~Kim}
\affiliation{Department of Physics, University of Toronto, Toronto, Ontario M5S 1A7, Canada}
\author{Young-June~Kim}
\affiliation{Department of Physics, University of Toronto, Toronto, Ontario M5S 1A7, Canada}
\author{N.~Doiron-Leyraud}
\affiliation{Institut Quantique, D\'epartement de physique \& RQMP, Universit\'e de Sherbrooke, Sherbrooke, Qu\'ebec J1K 2R1, Canada}
\author{L.~Taillefer}
\affiliation{Institut Quantique, D\'epartement de physique \& RQMP, Universit\'e de Sherbrooke, Sherbrooke, Qu\'ebec J1K 2R1, Canada}
\affiliation{Canadian Institute for Advanced Research, Toronto, Ontario M5G 1M1, Canada}


\date{\today}


\begin{abstract}

The material \RuCl~has been the subject of intense scrutiny as a potential Kitaev quantum spin liquid, predicted to display Majorana fermions as low energy excitations.
In practice, \RuCl~undergoes a transition to a state with antiferromagnetic order below a temperature $T_{\rm N}$ $\approx$ 7~K, but this order can be suppressed by applying an external in-plane magnetic field of $H_\parallel$ = 7~T.
Whether a quantum spin liquid phase exists just above that field is still an open question, but the reported observation of a quantized thermal Hall conductivity at $H_\parallel$ $>$ 7~T by Kasahara and co-workers $\big[$Kasahara {\it et al}., Nature {\bf 559}, 227 (2018)$\big]$ has been interpreted as evidence of itinerant Majorana fermions in the Kitaev quantum spin liquid state.
In this study, we re-examine the origin of the thermal Hall conductivity \Kxy~in \RuCl.
Our measurements of \Kxy($T$) on several different crystals yield a temperature dependence very similar to that of the phonon-dominated longitudinal thermal conductivity \Kxx($T$), for which the natural explanation is that \Kxy~is also mostly carried by phonons.
Upon cooling, \Kxx~peaks at $T \simeq$ 20~K, then drops until \TN, whereupon it suddenly increases again.
The abrupt increase below \TN~is attributed to a sudden reduction in the scattering of phonons by low-energy spin fluctuations as these become partially gapped when the system orders.
The fact that \Kxy~also increases suddenly below \TN~is strong evidence that the thermal Hall effect
in \RuCl~is also carried predominantly by phonons.
This implies that any quantized signal from Majorana edge modes would have to come on top 
of a sizable -- and sample-dependent -- phonon background.

\end{abstract}

\maketitle


\section{INTRODUCTION}

The quasi-2D Mott insulator \RuCl~has attracted much interest since it was proposed as a promising material for realizing a Kitaev spin liquid state, thanks to its honeycomb lattice with anisotropic bond-directional spin interactions~\cite{jackeli_mott_2009,plumb_$ensuremathalphaensuremath-mathrmrucl_3$:_2014}.
The Kitaev model predicts the existence of mobile Majorana fermions as quasiparticles that would manifest as topologically protected heat carriers on the edges of the sample.
In principle, such Majorana edge modes could be detected by measuring the thermal Hall effect~\cite{nasu_thermal_2017,ye_quantization_2018,vinkler-aviv_approximately_2018}, and their signature would be a half-integer quantized thermal Hall conductivity per plane $\kappa_{\rm xy}^{2 \rm D}$ = \Kxy $d$~at low temperature ($d$ is the honeycomb interlayer distance).
However, because \RuCl~orders antiferromagnetically at low temperature, below \TN~= 7~K,
in order to access the putative spin liquid phase one must apply an in-plane magnetic field in excess of a critical field $H_\parallel$ = 7~T to suppress the magnetic order.
The possibility thus arises that a spin liquid phase could emerge immediately above that critical field.
An inelastic neutron scattering study finds a magnetically disordered state at $T \to 0$, sandwiched between the ordered phase below $H_\parallel$ = 7~T and a field-polarized phase above $H_\parallel \simeq$ 9~T, which is reminiscent of a quantum spin liquid~\cite{balz_finite_2019}.
A few studies of heat transport have reported the observation of a quantized \Kxy~in some samples of \RuCl, in a narrow range of in-plane fields (6 $< H_\parallel <$ 9~T) and temperatures  (3 $< T < $ 6~K)~\cite{kasahara_majorana_2018,yokoi_half-integer_2021,bruin_robustness_2021}.
However, since then, a number of studies have reported \Kxy~signals in various insulators
and attributed them to phonons~\cite{grissonnanche_giant_2019,hirokane_phononic_2019,li_phonon_2020,grissonnanche_chiral_2020,boulanger_thermal_2020}.
The \Kxy~response can have either sign, and the magnitude of \Kxy~is easily as large as that found in \RuCl, often larger.
Typically, in those studies that attribute the thermal Hall effect to phonons, one finds that the magnitude of \Kxy~scales roughly with the magnitude of the phonon thermal conductivity \Kxx, with a ratio $|$\Kxy/\Kxx$|$ of order 10$^{-3}$.
These studies raise the possibility that phonons might also generate a thermal Hall effect in \RuCl.\\\\
In this paper, we explore that possibility.
We measured the thermal conductivity \Kxx~and thermal Hall conductivity \Kxy~in several samples of \RuCl, coming from two different sources.
We find a substantial variation in the magnitude of \Kxx~and \Kxy~from sample to sample, but the same qualitative behaviour.
Upon cooling from 80~K, \Kxx~and \Kxy~exhibit similar behaviours, namely a peak in $\kappa / T$ vs $T$ at roughly the same temperature $T \simeq$ 20~K, followed by a dip towards \TN.
If the magnetic field is applied along the $c$ direction, both \Kxx$/T$ and \Kxy$/T$ exhibit a rapid rise below \TN.
If a magnetic field is applied such that its in-plane component is $H_\parallel$ = 7~T, thereby suppressing the magnetic order, both \Kxx$/T$ and \Kxy$/T$ continue their monotonic decrease as $T \to 0$.
The fact that \Kxy($T$) mimics the phonon-dominated \Kxx($T$) is strong evidence that the thermal Hall signal is also dominated by phonons.
Moreover, we find that \Kxy/\Kxx~$\simeq$ 10$^{-3}$ at low temperature, a magnitude typical of the phonon Hall effect in other insulators.
Finally, the fact that there is a sizeable \Kxy~signal below \TN, in the antiferromagnetic phase, shows that it does not simply arise from the excitations of a pure spin liquid phase.
All this calls into question any interpretation of prior data exclusively in terms of Majorana fermions.



\section{METHODS}

{\it Samples}.~\textemdash~We measured a variety of samples coming from two different groups: Oak Ridge National Laboratory (ORNL) and the University of Toronto (UofT).
Single crystals from ORNL were grown using the vapor transport method on \RuCl~powder coming from Furuya Metal Co., Ltd; the growth technique is detailed elsewhere~\cite{may_practical_2020}.
Single crystals from UofT were grown using the same method but using powder coming from Sigma-Aldrich. 
The powder used was composed of 45\%-55\% ruthenium and was sealed in a quartz tube under vacuum. 
The latter was placed inside a two-zone tube furnace using a temperature gradient of 70$^\circ$ C (warmest side was 850$^\circ$ C). 
The powder was annealed over two days, followed by a 4$^\circ$ C/hour cooldown while maintaining the temperature gradient (see Ref.~\cite{sears_phase_2017} for more details).
Here we report data on four samples from ORNL (labelled O1, O2, O3 and O4) and one sample from UofT (labelled T1).
The samples were cut into thin rectangular platelets of typical dimensions 1 $\times$ 1~mm, with thicknesses ranging from 20 to 140 $\mu$m.
The contacts on the samples were made by glueing thin silver leads with silver paste.
%


{\it Measurement technique}.~\textemdash~Measurements were performed by a steady-state method using a standard four-terminal technique, with the thermal current applied along the length $l$ of the sample within the honeycomb layers (perpendicular to the Ru-Ru bonds; $J||a$).
The thermal conductivity \Kxx~was measured by employing a standard one-heater--two-thermometers method, using a 5 k$\rm \Omega$ resistor and two {\it in situ} calibrated bare chip CX-1050 Cernox sensors.
A constant heat current $\dot{Q}$ was injected at one end of the sample, while at the other end the sample is well heat sunk to a copper block referenced at a temperature $T_0$.
The heat current was generated by sending an electrical current through a 5 k$\rm \Omega$ strain gauge whose resistance is marginally dependent of temperature and magnetic field.
A longitudinal thermal gradient $\Delta T_{\rm x} = T^+ - T^-$ is measured at two points along the length of the sample, separated by a distance $l$.
The longitudinal thermal conductivity is given by $\kappa_{\rm xx} = \dot{Q} / \left( \Delta T_{\rm x} \alpha \right)$, where $\alpha = w t / l$ is the geometric factor of the sample (width $w$, thickness $t$).
Under a magnetic field applied perpendicular to the basal plane, a transverse thermal gradient $\Delta T_{\rm y}$ resulted (the field can also be tilted at an angle, such that it will have a component in the plane ($H_\parallel$), as well as normal to the plane ($H_\perp$)).
This transverse gradient was measured using a differential type-E thermocouple.
The thermal Hall conductivity is then given by $\kappa_{\rm xy} = -\kappa_{\rm yy} \left( \Delta T_{\rm y} / \Delta T_{\rm x} \right) \left( l / w \right)$ after having antisymmetrized the thermal Hall gradient via $\Delta T_{\rm y}\left( H \right) = \left[\Delta T_{\rm y}\left(T, H\right) - \Delta T_{\rm y}\left(T, -H\right)\right] / 2$.
The error bars on the absolute values of thermal coefficients come mostly from the uncertainty in estimating the dimensions ($l$, $w$ and $t$) of the samples, approximately $\pm$20\%.
The applied current was chosen such that $\Delta T / T$ $\simeq$ 5-10\%; the resulting \Kxx~was independent of $\Delta T$, indicating that there was no heat loss.
Any contamination of $\Delta T_{\rm y}(H)$ coming from the copper heat sink was ruled out by a previous study that compared heat sinks made of Cu vs LiF (see Supplementary Information in Ref.~\onlinecite{boulanger_thermal_2020}).
Moreover, $\Delta T_{\rm y}$ data obtained with thermocouples were found to be in good agreement with $\Delta T_{\rm y}$ data obtained with Cernox sensors applied to the same sample.


\section{RESULTS}

\begin{figure}[t]\centering
\includegraphics[width = 0.4\textwidth]{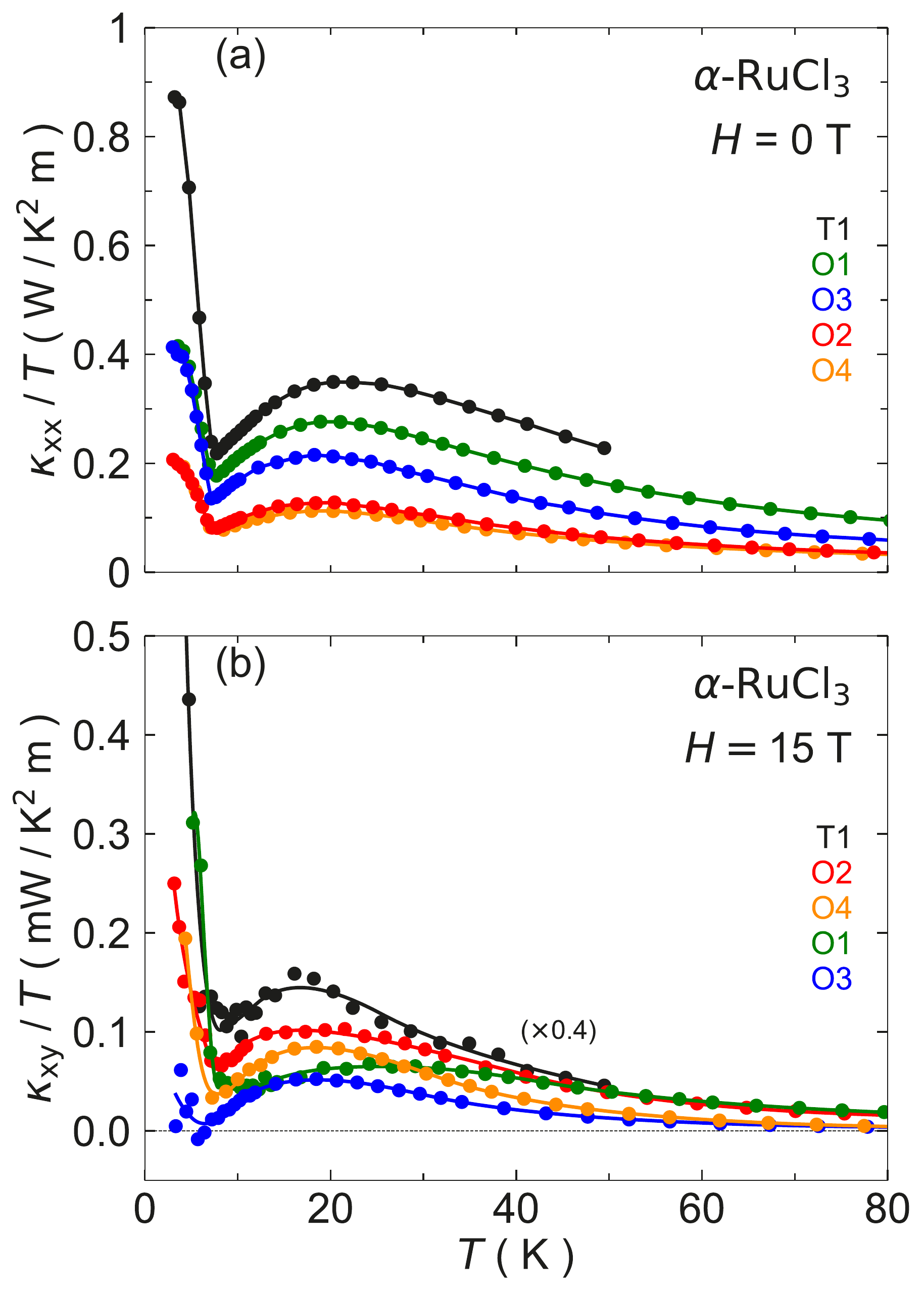}
\caption{
(a) Thermal conductivity of \RuCl~in zero magnetic field, plotted as \Kxx$/T$ vs $T$, for our five different samples: T1 (black), O1 (green), O2 (red), O3 (blue) and O4 (gold). 
(b) Thermal Hall conductivity of the same five samples, measured in a magnetic field of $H$ = 15~T applied normal to the honeycomb layers ($ab$-plane), plotted as \Kxy$/T$ vs $T$. 
The \Kxy~data for sample T1 (black) have been multiplied by a factor 0.4.
Lines are a guide to the eye.}
\label{fig1}
\end{figure}

In Fig.~\ref{fig1}(a), we show the thermal conductivity of \newline\RuCl~measured in five crystals, plotted as \Kxx$/T$ vs $T$.
We see that there is a considerable variation in the magnitude of \Kxx~amongst samples, by a factor 3 or so, with the largest \Kxx$(T)$ value seen in sample T1.
Prior data by Leahy \etal~\cite{leahy_anomalous_2017} and Hentrich \etal~\cite{hentrich_unusual_2018} fall within the range of magnitudes of our own samples.
Kasahara \etal~\cite{kasahara_unusual_2018} find a \Kxx$(T)$ that is 2-3 (7-8) times larger than our data on sample T1 (O2).
This variation in magnitude is attributed to different levels of disorder, perhaps structural (associated with domains that form upon cooling through the structural transition at 130~K).
Despite this quantitative variation, the qualitative behavior of \Kxx($T$) is the same in all samples.
There is a peak in \Kxx$/T$ vs $T$ at $T \simeq$ 20~K, below which \Kxx$/T$ drops as $T \to$ \TN.
As argued by Hentrich \etal~\cite{hentrich_unusual_2018}, the dominant carriers of heat in \RuCl~are phonons, and these become increasingly scattered by low-energy antiferromagnetic spin fluctuations upon cooling below 80~K.
It is this scattering that causes \Kxx$/T$ to drop as $T \to$~\TN~(from above).
Application of a magnetic field in the plane gaps the low-energy spin fluctuation spectrum, causing \Kxx~at $T$ = 10~K ($>$~\TN) to increase rapidly for $H >$ 10~T~\cite{hentrich_unusual_2018}.\\\\
As we cool below \TN, at $T <$ 1~K ($<$~\TN), we see that \Kxx$/T$ (in zero field) shoots up immediately in all samples (Fig.~\ref{fig1}a and Refs.~\cite{leahy_anomalous_2017,hentrich_unusual_2018,kasahara_unusual_2018}), presumably because the magnetic order also causes a gapping of the low-energy spin fluctuation spectrum.
This V-shaped dependence of \Kxx$/T$ vs T at $H$ = 0~T is mimicked by a similar V-shaped dependence of \Kxx~vs $H$ at $T <$ 1~K, with the minimum at the critical field of 7~T~\cite{yu_ultralow-temperature_2018}.
In summary, the thermal conductivity of \RuCl~at low temperature ($T <$ 50~K) can be understood essentially in terms of phonons scattered by spin fluctuations.
%


In Fig.~\ref{fig1}(b), we show the thermal Hall conductivity of \RuCl~measured in the same 5 samples, plotted as \Kxy$/T$ vs $T$ for a field $H$ = 15~T applied normal to the plane ($H||c$).
There is considerable variation in the magnitude of \Kxy~across samples, even larger than that seen in \Kxx.
Prior data by Hentrich \etal~\cite{hentrich_large_2019} (with $H$ = 16~T) yield a \Kxy($T$) curve very similar, quantitatively and qualitatively, to the curve for our sample O2 (for which $H$ = 15~T).
Kasahara \etal~\cite{kasahara_unusual_2018} find a \Kxy$(T)$ curve that is 3-4 times larger than that.
Despite the quantitative variation, the qualitative behavior of \Kxy($T$) is the same in all samples from all groups -- at least for $T >$ \TN.
The thermal Hall signal is positive, and there is a peak in \Kxy$/T$ vs $T$ at $T \simeq$ 20~K, below which \Kxy$/T$ drops as $T \to$ \TN.
In other words, \Kxy$(T)$ mimics the phonon-dominated \Kxx$(T)$.
%


\begin{figure}[t]\centering
\includegraphics[width = 0.35\textwidth]{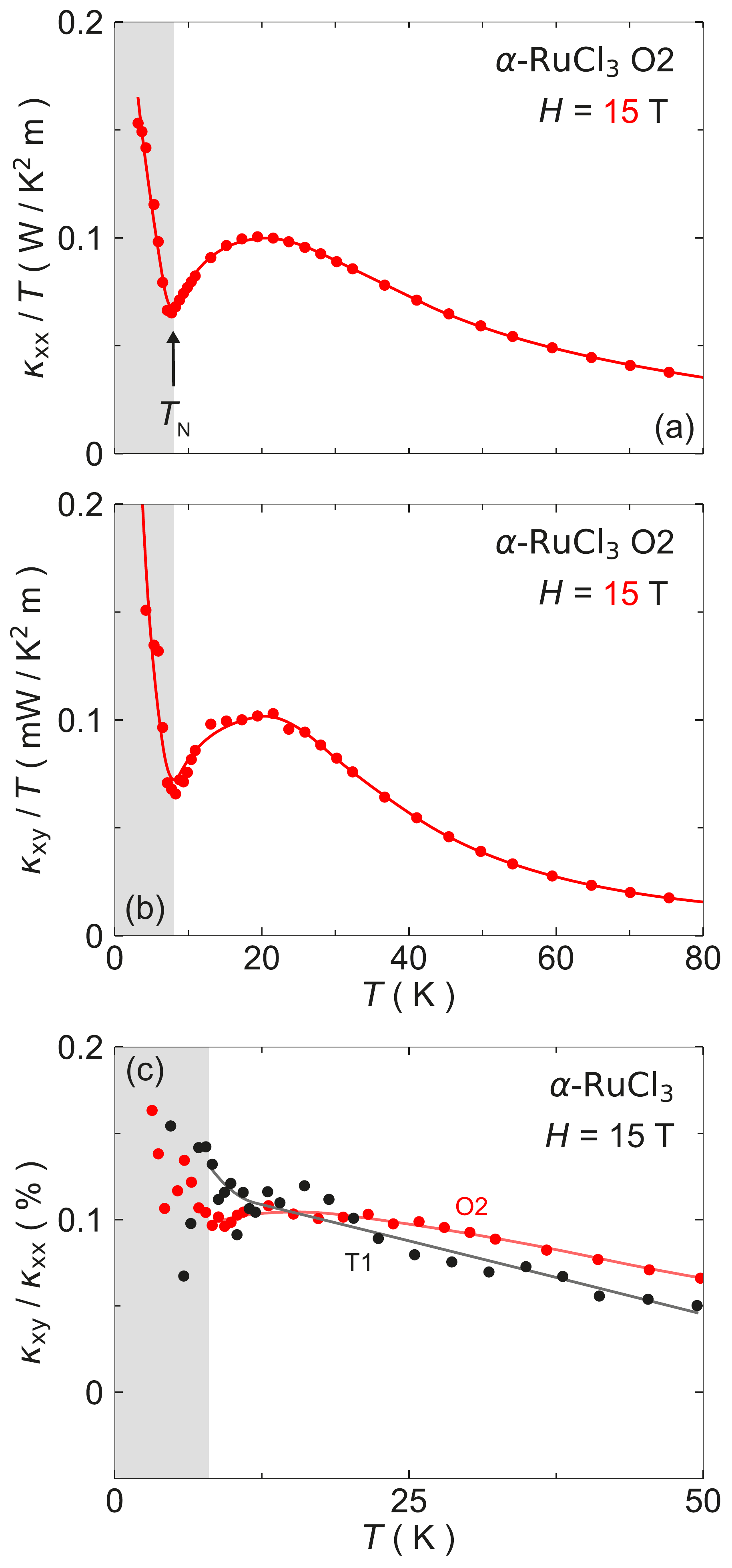}
\caption{
(a) Thermal conductivity of sample O2, plotted as \Kxx$/T$ vs $T$, for $H || c$ = 15~T.
(b) Thermal Hall conductivity of the same sample, plotted as \Kxy$/T$ vs $T$, under the same conditions. 
(c) Ratio \Kxy/\Kxx~(sample O2, red): data in (b) over data in (a). Corresponding data for sample T1 are shown in black.
The gray shaded area delineates the region below \TN(arrow) where antiferromagnetic order is present. 
Lines are a guide to the eye.}
\label{fig2}
\end{figure}

\begin{figure}[t]\centering
\includegraphics[width = 0.4\textwidth]{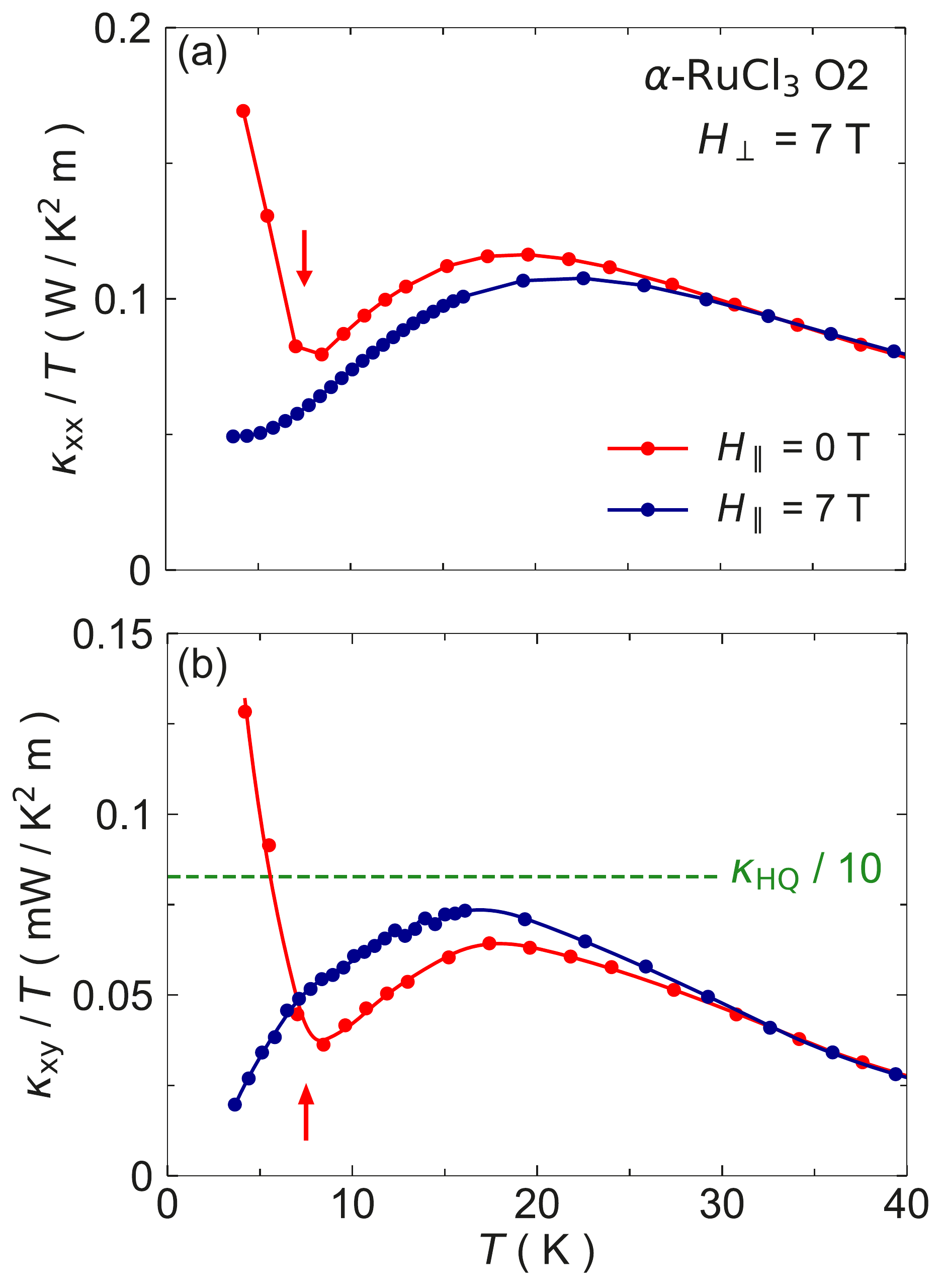}
\caption{
(a) Thermal conductivity \Kxx~of sample O2, plotted as \Kxx$/T$ vs $T$, for an applied magnetic whose component normal to the honeycomb layers is $H_{\perp}$ = 7~T and whose component parallel to the layers is either $H_\parallel$ = 0~T (red) or $H_\parallel$ = 7~T (blue).
(b) Same as in (a), for \Kxy.
The horizontal dashed line marks the quantized value ($\kappa_{\rm HQ}$) expected for Majorana edge modes, divided by 10.
The arrows mark \TN.
Lines are a guide to the eye.}
\label{fig3}
\end{figure}

Below \TN, we observe an increase in \Kxy$(T)$ upon cooling (Fig.~\ref{fig1}(b)), in all our samples.
Although \Kxy$(T)$ dips down to a minimum at \TN, in one case getting close to zero (sample O4), it always remains positive. 
Data by Hentrich \etal~\cite{hentrich_large_2019} on one sample show a  \Kxy$(T)$ curve that dips to zero at \TN, becoming perhaps very slightly negative; however, no data were reported for $T<$ 7~K.
The data by Kasahara \etal~\cite{kasahara_unusual_2018} do extend below \TN, where they report a negative \Kxy~signal, in contrast to our positive signal.
It is not clear where this discrepancy comes from.
We note that Kasahara \etal also report that \Kxy~goes from positive to negative when the temperature is raised across $T \simeq$ 55~K, a sign change that is seen neither by Hentrich \etal~nor by us, in any sample (Fig.~\ref{fig1}(b)).
In summary, our \Kxy($T$) data from all five samples show a clear similarity with the phonon-dominated \Kxx($T$) data, not only above \TN, but also below \TN. 
In Fig.~\ref{fig2}, we compare these two curves for one sample (O2), where the similarity between \Kxx($T$) (panel a) and \Kxy($T$) (panel b) is striking.
In panel (c), we plot the ratio \Kxy/\Kxx~vs $T$, measured at $H$ = 15~T, and see that \Kxy~is approximately 1000 smaller than \Kxx, at $T \simeq$ 20~K.
The same is true for sample T1, whose \Kxy~amplitude is 3-4 times larger (at $T \simeq$ 20~K; Fig.~\ref{fig1}).
For our five samples, the ratio at $T$ = 20~K and $H$ = 15~T ranges from 0.03 \% to 0.10 \%.
This is consistent with prior data by Hentrich \etal~\cite{hentrich_large_2019} and Kasahara \etal~\cite{kasahara_unusual_2018}, where \Kxy/\Kxx~$\simeq$ 0.05~\%.
As we will discuss below, a ratio of this magnitude is typical for a phonon thermal Hall effect.
In Fig.~\ref{fig3}, we show the effect of applying a component of the field parallel to the honeycomb layers, so that the antiferromagnetic order is suppressed.
The red curves are for $H_{\perp}$ = 7~T and no in-plane field ($H_\parallel$ = 0~T).
Both curves --  \Kxx($T$) (panel a) and \Kxy($T$) (panel b) -- are very similar to those in Fig.~\ref{fig2}, where $H_{\perp}$ = 15~T (and $H_\parallel$ = 0~T).
The only difference is quantitative: the magnitude of \Kxy~is down roughly by a factor 7 T / 15 T (the reduction in $H_\perp$).
Adding an additional 7~T field component in the plane (blue curves) has only a small effect above 7~K, but a dramatic one below 7~K: now \Kxx($T$) and \Kxy($T$) no longer suddenly increase below 7~K (= \TN) but continue to decrease smoothly through 7~K, in the absence of ordering.
This makes sense for phonons, which continue to be scattered by low-lying spin fluctuations that remain ungapped down to the lowest temperature.
So again, the striking similarity between \Kxy($T$) and the phonon-dominated \Kxx($T$) argues for a thermal Hall signal carried by the phonons.
%


\section{DISCUSSION}

Traditionally, the thermal Hall effect from phonons has been considered very small~\cite{hirschberger_thermal_2015}.
First observed in 2005, its magnitude in Tb$_3$Ga$_5$O$_{12}$ was detected to be \Kxy~$\simeq 0.02$~mW/K m at $H$ = 3~T and $T$ = 5~K~\cite{strohm_phenomenological_2005}.
This is two orders of magnitude smaller than the signal first reported in \RuCl, in 2017: \Kxy~$\simeq$ 4~mW/K m at $H$ = 6~T and $T$ = 20~K~\cite{kasahara_unusual_2018}.
So it was perhaps natural to rule out phonons in 2017, although that year a large thermal Hall effect was reported in the antiferromagnetic insulator Fe$_2$Mo$_3$O$_8$ and attributed to phonons (and their coupling to spins), with \Kxy~$\simeq$ 30~mW/K m at $H$ = 14~T and $T$ = 35~K~\cite{ideue_giant_2017}, one order of magnitude larger than in \RuCl.
Since then, it has become abundantly clear that phonons can carry a sizable thermal Hall conductivity in various insulators, whether magnetic, as in the antiferromagnetic cuprate Mott insulator La$_2$CuO$_4$~\cite{grissonnanche_giant_2019,grissonnanche_chiral_2020}, or non-magnetic, as in the quantum paraelectric SrTiO$_3$~\cite{li_phonon_2020}.
The magnitude of \Kxy~(at $H$ = 15~T and $T$ = 20~K) can vary from $|\kappa_{xy}| \simeq$ 1~mW/K m in Tb$_2$Ti$_2$O$_7$~\cite{hirschberger_thermal_2015,hirokane_phononic_2019} to $|\kappa_{xy}| \simeq$ 200~mW/K m in Nd$_2$CuO$_4$~\cite{boulanger_thermal_2020} and become as large as $|\kappa_{xy}| \simeq$ 1000~mW/K m in Cu$_3$TeO$_6$~\cite{chen_large_2021}.
Although the mechanisms by which phonons acquire a handedness (become chiral) in a magnetic field are still unclear in these materials, the cumulative evidence that phonons are the carriers of the thermal Hall effect in these insulators is strong: although $|\kappa_{xy}|$ varies by three orders of magnitude, the ratio of \Kxy~over the phonon-dominated \Kxx~is roughly the same, namely $|\kappa_{xy} / \kappa_{xx}| \simeq 2-5 \times 10^{-3}$ in all cases (see Table 1 in~\cite{chen_large_2021}).
In other words, what really varies from material to material is the ability of phonons to conduct heat, \ie~the magnitude of \Kxx.
Being now aware of these more recent studies, it seems very likely that phonons must also generate a sizable \Kxy~signal in the antiferromagnetic insulator \RuCl, especially given the measured ratio $|\kappa_{xy} / \kappa_{xx}| \simeq 1 \times 10^{-3}$ (at $H$ = 15~T and $T$ = 20~K; Fig.~\ref{fig2}(b)).
The immediate implication of having a sizable phonon contribution to \Kxy~in \RuCl~is that the total measured value of \Kxy~in any sample would include a sizable phonon background, to which any contribution from Majorana fermions would add.
In this context, the reported observation of a plateau in \Kxy$/T$, over a small range of in-plane fields (6 $< H_\parallel <$ 9~T) and temperatures (3 $< T <$ 6~K)~\cite{kasahara_majorana_2018}, with a measured total value of \Kxy$/T$ having precisely the half-quantized value of $\pi k_{\rm B}^2/12 \hbar$ per plane expected theoretically for Majorana edge modes~\cite{nasu_thermal_2017}, can only be meaningful if the phonon background is precisely zero.
As we have argued, this condition is unlikely to be satisfied in \RuCl, at least when there is a component of the magnetic field normal to the plane (and to the heat current), which was the case in the study of Ref.~\onlinecite{kasahara_majorana_2018}.
To investigate the same configuration as in Ref.~\onlinecite{kasahara_majorana_2018}, we turn to our data in a tilted field, displayed in Fig.~\ref{fig3} (blue curves).
If the entire signal is attributed to Majorana fermions, it is difficult to understand why our value of \Kxy$/T$ at $H_\parallel$ = 7~T and $T$ = 6~K is only \Kxy$/T$ $\simeq$ 0.04~mW/K$^2$ m (Fig.~\ref{fig3}b), while the value reported in Ref.~\onlinecite{kasahara_majorana_2018} is \Kxy$/T$ $\simeq$ 0.8~mW/K$^2$ m, a factor 20 larger.
On the other hand, if the signal is attributed to phonons, then the likely reason for the much smaller \Kxy~in our sample is simply that its \Kxx~is much smaller, with \Kxx~$\simeq$ 0.5~W/Km at $T$ = 10~K and $H_\parallel$ = 7~T (Fig.~\ref{fig3}a), while the value reported in Ref.~\onlinecite{kasahara_majorana_2018} is $\kappa_{xx} \simeq 5$~W/Km -- a factor of 10 larger.
In other words, a phonon scenario solves the puzzle of why a thermal Hall conductivity that exceeds the half-quantized value, \Kxy$/T$ = $\kappa_{\rm HQ}$ = 0.8 mW/K$^2$ m has only been observed in samples that have the largest values of \Kxx~\cite{kasahara_majorana_2018,yokoi_half-integer_2021,bruin_robustness_2021,yamashita_sample_2020}: a larger value of \Kxy~is what we expect from phonons that conduct better (and thus produce a larger \Kxx), as nicely demonstrated by the highly conductive samples of Cu$_3$TeO$_6$~\cite{chen_large_2021}.


\section{SUMMARY}

We have measured the thermal conductivity \Kxx~and the thermal Hall conductivity \Kxy~of \RuCl, for a heat current within the honeycomb layers (perpendicular to the Ru-Ru bond; $J||a$) in five different single crystals, from two separate sources.
Although the magnitude of \Kxx~and \Kxy~vary significantly from sample to sample, presumably because of varying degrees of structural disorder, we find the same qualitative behavior in all samples.
Upon cooling, both \Kxx$(T)$ and \Kxy$(T)$ are found to have the same temperature dependence, namely a broad peak located at the same temperature $T \simeq$ 20~K, followed by a decrease until $T$ = 7~K, whereupon \Kxx$(T)$ and \Kxy$(T)$ either rapidly rise in tandem upon entering the antiferromagnetic phase at \TN~= 7~K, when the field is applied normal to the layers, or both continue their decrease, when the field has an in-plane component sufficient to remove the antiferromagnetic order.
The fact that  \Kxy$(T)$ mimics the phonon-dominated \Kxx$(T)$ so well leads us to conclude that the thermal Hall effect in \RuCl~is carried predominantly by phonons.
This interpretation is supported by the fact that the magnitude of \Kxy~in various samples of \RuCl~roughly scales with the magnitude of the phonon-dominated \Kxx.
Moreover, the ratio \Kxy/\Kxx~has a magnitude comparable to that found in several other insulators where phonons have been shown or argued to cause the Hall effect, namely $|\kappa_{xy} / \kappa_{xx}| \simeq 1 \times 10^{-3}$ (see Table 1 in~\cite{chen_large_2021}).
If phonons contribute significantly to the \Kxy~signal of \RuCl~samples, the experimentally measured \Kxy~cannot be attributed directly and exclusively to theoretically predicted Majorana fermions, as has been done in some reports.


\section*{ACKNOWLEDGMENTS}

We thank S.~Fortier for his assistance with the experiments and M.~Dion for his assistance with sample orientation.
L.T. acknowledges support from the Canadian Institute for Advanced Research (CIFAR) as a CIFAR Fellow and funding from the Institut Quantique, the Natural Sciences and Engineering Research Council of Canada (NSERC; PIN:123817), the Fonds de Recherche du Qu\'{e}bec -- Nature et Technologies (FRQNT), the Canada Foundation for Innovation (CFI), and a Canada Research Chair.
This research was undertaken thanks in part to funding from the Canada First Research Excellence Fund.
C.B. and S.E.N. were supported by the U.S. Department of Energy (DOE), Basic Energy Sciences, Scientific User Facilities Division.
S.E.N. also acknowledges support from the Quantum Science Center (QSC), a National Quantum Information Science Research Center of the U.S. DOE.
Materials synthesis by J.Q.Y. was supported by the US DOE, Office of Science, Basic Energy Sciences, Materials Sciences and Engineering Division.
D.G.M. acknowledges support from the Gordon and Betty Moore Foundation’s EPiQS Initiative, Grant GBMF9069.
Work at the University of Toronto was supported by NSERC, CFI, and Ontario Research Fund.

%

\begin{thebibliography}{26}%
    \makeatletter
    \providecommand \@ifxundefined [1]{%
     \@ifx{#1\undefined}
    }%
    \providecommand \@ifnum [1]{%
     \ifnum #1\expandafter \@firstoftwo
     \else \expandafter \@secondoftwo
     \fi
    }%
    \providecommand \@ifx [1]{%
     \ifx #1\expandafter \@firstoftwo
     \else \expandafter \@secondoftwo
     \fi
    }%
    \providecommand \natexlab [1]{#1}%
    \providecommand \enquote  [1]{``#1''}%
    \providecommand \bibnamefont  [1]{#1}%
    \providecommand \bibfnamefont [1]{#1}%
    \providecommand \citenamefont [1]{#1}%
    \providecommand \href@noop [0]{\@secondoftwo}%
    \providecommand \href [0]{\begingroup \@sanitize@url \@href}%
    \providecommand \@href[1]{\@@startlink{#1}\@@href}%
    \providecommand \@@href[1]{\endgroup#1\@@endlink}%
    \providecommand \@sanitize@url [0]{\catcode `\\12\catcode `\$12\catcode
      `\&12\catcode `\#12\catcode `\^12\catcode `\_12\catcode `\%12\relax}%
    \providecommand \@@startlink[1]{}%
    \providecommand \@@endlink[0]{}%
    \providecommand \url  [0]{\begingroup\@sanitize@url \@url }%
    \providecommand \@url [1]{\endgroup\@href {#1}{\urlprefix }}%
    \providecommand \urlprefix  [0]{URL }%
    \providecommand \Eprint [0]{\href }%
    \providecommand \doibase [0]{https://doi.org/}%
    \providecommand \selectlanguage [0]{\@gobble}%
    \providecommand \bibinfo  [0]{\@secondoftwo}%
    \providecommand \bibfield  [0]{\@secondoftwo}%
    \providecommand \translation [1]{[#1]}%
    \providecommand \BibitemOpen [0]{}%
    \providecommand \bibitemStop [0]{}%
    \providecommand \bibitemNoStop [0]{.\EOS\space}%
    \providecommand \EOS [0]{\spacefactor3000\relax}%
    \providecommand \BibitemShut  [1]{\csname bibitem#1\endcsname}%
    \let\auto@bib@innerbib\@empty
    \bibitem [{\citenamefont {Jackeli}\ and\ \citenamefont
      {Khaliullin}(2009)}]{jackeli_mott_2009}%
      \BibitemOpen
      \bibfield  {author} {\bibinfo {author} {\bibfnamefont {G.}~\bibnamefont
      {Jackeli}}\ and\ \bibinfo {author} {\bibfnamefont {G.}~\bibnamefont
      {Khaliullin}},\ }\bibfield  {title} {\bibinfo {title} {{Mott} {Insulators} in
      the {Strong} {Spin}-{Orbit} {Coupling} {Limit}: {From} {Heisenberg} to a
      {Quantum} {Compass} and {Kitaev} {Models}},\ }\href
      {https://doi.org/10.1103/PhysRevLett.102.017205} {\bibfield  {journal}
      {\bibinfo  {journal} {Physical Review Letters}\ }\textbf {\bibinfo {volume}
      {102}},\ \bibinfo {pages} {017205} (\bibinfo {year} {2009})}\BibitemShut
      {NoStop}%
    \bibitem [{\citenamefont {Plumb}\ \emph {et~al.}(2014)\citenamefont {Plumb},
      \citenamefont {Clancy}, \citenamefont {Sandilands}, \citenamefont {Shankar},
      \citenamefont {Hu}, \citenamefont {Burch}, \citenamefont {Kee},\ and\
      \citenamefont {Kim}}]{plumb_$ensuremathalphaensuremath-mathrmrucl_3$:_2014}%
      \BibitemOpen
      \bibfield  {author} {\bibinfo {author} {\bibfnamefont {K.~W.}\ \bibnamefont
      {Plumb}}, \bibinfo {author} {\bibfnamefont {J.~P.}\ \bibnamefont {Clancy}},
      \bibinfo {author} {\bibfnamefont {L.~J.}\ \bibnamefont {Sandilands}},
      \bibinfo {author} {\bibfnamefont {V.~V.}\ \bibnamefont {Shankar}}, \bibinfo
      {author} {\bibfnamefont {Y.~F.}\ \bibnamefont {Hu}}, \bibinfo {author}
      {\bibfnamefont {K.~S.}\ \bibnamefont {Burch}}, \bibinfo {author}
      {\bibfnamefont {H.-Y.}\ \bibnamefont {Kee}},\ and\ \bibinfo {author}
      {\bibfnamefont {Y.-J.}\ \bibnamefont {Kim}},\ }\bibfield  {title} {\bibinfo
      {title} {$\alpha$-{RuCl}$_3$: {A} spin-orbit assisted {Mott} insulator on a
      honeycomb lattice},\ }\href {https://doi.org/10.1103/PhysRevB.90.041112}
      {\bibfield  {journal} {\bibinfo  {journal} {Physical Review B}\ }\textbf
      {\bibinfo {volume} {90}},\ \bibinfo {pages} {041112} (\bibinfo {year}
      {2014})}\BibitemShut {NoStop}%
    \bibitem [{\citenamefont {Nasu}\ \emph {et~al.}(2017)\citenamefont {Nasu},
      \citenamefont {Yoshitake},\ and\ \citenamefont {Motome}}]{nasu_thermal_2017}%
      \BibitemOpen
      \bibfield  {author} {\bibinfo {author} {\bibfnamefont {J.}~\bibnamefont
      {Nasu}}, \bibinfo {author} {\bibfnamefont {J.}~\bibnamefont {Yoshitake}},\
      and\ \bibinfo {author} {\bibfnamefont {Y.}~\bibnamefont {Motome}},\
      }\bibfield  {title} {\bibinfo {title} {{Thermal} {Transport} in the {Kitaev}
      {Model}},\ }\href {https://doi.org/10.1103/PhysRevLett.119.127204} {\bibfield
       {journal} {\bibinfo  {journal} {Physical Review Letters}\ }\textbf {\bibinfo
      {volume} {119}},\ \bibinfo {pages} {127204} (\bibinfo {year}
      {2017})}\BibitemShut {NoStop}%
    \bibitem [{\citenamefont {Ye}\ \emph {et~al.}(2018)\citenamefont {Ye},
      \citenamefont {Hal\'asz}, \citenamefont {Savary},\ and\ \citenamefont
      {Balents}}]{ye_quantization_2018}%
      \BibitemOpen
      \bibfield  {author} {\bibinfo {author} {\bibfnamefont {M.}~\bibnamefont
      {Ye}}, \bibinfo {author} {\bibfnamefont {G.~B.}\ \bibnamefont {Hal\'asz}},
      \bibinfo {author} {\bibfnamefont {L.}~\bibnamefont {Savary}},\ and\ \bibinfo
      {author} {\bibfnamefont {L.}~\bibnamefont {Balents}},\ }\bibfield  {title}
      {\bibinfo {title} {{Quantization} of the {Thermal} {Hall} {Conductivity} at
      {Small} {Hall} {Angles}},\ }\href
      {https://doi.org/10.1103/PhysRevLett.121.147201} {\bibfield  {journal}
      {\bibinfo  {journal} {Physical Review Letters}\ }\textbf {\bibinfo {volume}
      {121}},\ \bibinfo {pages} {147201} (\bibinfo {year} {2018})}\BibitemShut
      {NoStop}%
    \bibitem [{\citenamefont {Vinkler-Aviv}\ and\ \citenamefont
      {Rosch}(2018)}]{vinkler-aviv_approximately_2018}%
      \BibitemOpen
      \bibfield  {author} {\bibinfo {author} {\bibfnamefont {Y.}~\bibnamefont
      {Vinkler-Aviv}}\ and\ \bibinfo {author} {\bibfnamefont {A.}~\bibnamefont
      {Rosch}},\ }\bibfield  {title} {\bibinfo {title} {{Approximately} {Quantized}
      {Thermal} {Hall} {Effect} of {Chiral} {Liquids} {Coupled} to {Phonons}},\
      }\href {https://doi.org/10.1103/PhysRevX.8.031032} {\bibfield  {journal}
      {\bibinfo  {journal} {Physical Review X}\ }\textbf {\bibinfo {volume} {8}},\
      \bibinfo {pages} {031032} (\bibinfo {year} {2018})}\BibitemShut {NoStop}%
    \bibitem [{\citenamefont {Balz}\ \emph {et~al.}(2019)\citenamefont {Balz},
      \citenamefont {Lampen-Kelley}, \citenamefont {Banerjee}, \citenamefont {Yan},
      \citenamefont {Lu}, \citenamefont {Hu}, \citenamefont {Yadav}, \citenamefont
      {Takano}, \citenamefont {Liu}, \citenamefont {Tennant}, \citenamefont
      {Lumsden}, \citenamefont {Mandrus},\ and\ \citenamefont
      {Nagler}}]{balz_finite_2019}%
      \BibitemOpen
      \bibfield  {author} {\bibinfo {author} {\bibfnamefont {C.}~\bibnamefont
      {Balz}}, \bibinfo {author} {\bibfnamefont {P.}~\bibnamefont {Lampen-Kelley}},
      \bibinfo {author} {\bibfnamefont {A.}~\bibnamefont {Banerjee}}, \bibinfo
      {author} {\bibfnamefont {J.}~\bibnamefont {Yan}}, \bibinfo {author}
      {\bibfnamefont {Z.}~\bibnamefont {Lu}}, \bibinfo {author} {\bibfnamefont
      {X.}~\bibnamefont {Hu}}, \bibinfo {author} {\bibfnamefont {S.~M.}\
      \bibnamefont {Yadav}}, \bibinfo {author} {\bibfnamefont {Y.}~\bibnamefont
      {Takano}}, \bibinfo {author} {\bibfnamefont {Y.}~\bibnamefont {Liu}},
      \bibinfo {author} {\bibfnamefont {D.~A.}\ \bibnamefont {Tennant}}, \bibinfo
      {author} {\bibfnamefont {M.~D.}\ \bibnamefont {Lumsden}}, \bibinfo {author}
      {\bibfnamefont {D.}~\bibnamefont {Mandrus}},\ and\ \bibinfo {author}
      {\bibfnamefont {S.~E.}\ \bibnamefont {Nagler}},\ }\bibfield  {title}
      {\bibinfo {title} {{Finite} field regime for a quantum spin liquid in
      $\alpha$-{RuCl}$_3$},\ }\href {https://doi.org/10.1103/PhysRevB.100.060405}
      {\bibfield  {journal} {\bibinfo  {journal} {Physical Review B}\ }\textbf
      {\bibinfo {volume} {100}},\ \bibinfo {pages} {060405} (\bibinfo {year}
      {2019})}\BibitemShut {NoStop}%
    \bibitem [{\citenamefont {Kasahara}\ \emph
      {et~al.}(2018{\natexlab{a}})\citenamefont {Kasahara}, \citenamefont
      {Ohnishi}, \citenamefont {Mizukami}, \citenamefont {Tanaka}, \citenamefont
      {Ma}, \citenamefont {Sugii}, \citenamefont {Kurita}, \citenamefont {Tanaka},
      \citenamefont {Nasu}, \citenamefont {Motome}, \citenamefont {Shibauchi},\
      and\ \citenamefont {Matsuda}}]{kasahara_majorana_2018}%
      \BibitemOpen
      \bibfield  {author} {\bibinfo {author} {\bibfnamefont {Y.}~\bibnamefont
      {Kasahara}}, \bibinfo {author} {\bibfnamefont {T.}~\bibnamefont {Ohnishi}},
      \bibinfo {author} {\bibfnamefont {Y.}~\bibnamefont {Mizukami}}, \bibinfo
      {author} {\bibfnamefont {O.}~\bibnamefont {Tanaka}}, \bibinfo {author}
      {\bibfnamefont {S.}~\bibnamefont {Ma}}, \bibinfo {author} {\bibfnamefont
      {K.}~\bibnamefont {Sugii}}, \bibinfo {author} {\bibfnamefont
      {N.}~\bibnamefont {Kurita}}, \bibinfo {author} {\bibfnamefont
      {H.}~\bibnamefont {Tanaka}}, \bibinfo {author} {\bibfnamefont
      {J.}~\bibnamefont {Nasu}}, \bibinfo {author} {\bibfnamefont {Y.}~\bibnamefont
      {Motome}}, \bibinfo {author} {\bibfnamefont {T.}~\bibnamefont {Shibauchi}},\
      and\ \bibinfo {author} {\bibfnamefont {Y.}~\bibnamefont {Matsuda}},\
      }\bibfield  {title} {\bibinfo {title} {{Majorana} quantization and
      half-integer thermal quantum {Hall} effect in a {Kitaev} spin liquid},\
      }\href {https://doi.org/10.1038/s41586-018-0274-0} {\bibfield  {journal}
      {\bibinfo  {journal} {Nature}\ }\textbf {\bibinfo {volume} {559}},\ \bibinfo
      {pages} {227} (\bibinfo {year} {2018}{\natexlab{a}})}\BibitemShut {NoStop}%
    \bibitem [{\citenamefont {Yokoi}\ \emph {et~al.}(2021)\citenamefont {Yokoi},
      \citenamefont {Ma}, \citenamefont {Kasahara}, \citenamefont {Kasahara},
      \citenamefont {Shibauchi}, \citenamefont {Kurita}, \citenamefont {Tanaka},
      \citenamefont {Nasu}, \citenamefont {Motome}, \citenamefont {Hickey},
      \citenamefont {Trebst},\ and\ \citenamefont
      {Matsuda}}]{yokoi_half-integer_2021}%
      \BibitemOpen
      \bibfield  {author} {\bibinfo {author} {\bibfnamefont {T.}~\bibnamefont
      {Yokoi}}, \bibinfo {author} {\bibfnamefont {S.}~\bibnamefont {Ma}}, \bibinfo
      {author} {\bibfnamefont {Y.}~\bibnamefont {Kasahara}}, \bibinfo {author}
      {\bibfnamefont {S.}~\bibnamefont {Kasahara}}, \bibinfo {author}
      {\bibfnamefont {T.}~\bibnamefont {Shibauchi}}, \bibinfo {author}
      {\bibfnamefont {N.}~\bibnamefont {Kurita}}, \bibinfo {author} {\bibfnamefont
      {H.}~\bibnamefont {Tanaka}}, \bibinfo {author} {\bibfnamefont
      {J.}~\bibnamefont {Nasu}}, \bibinfo {author} {\bibfnamefont {Y.}~\bibnamefont
      {Motome}}, \bibinfo {author} {\bibfnamefont {C.}~\bibnamefont {Hickey}},
      \bibinfo {author} {\bibfnamefont {S.}~\bibnamefont {Trebst}},\ and\ \bibinfo
      {author} {\bibfnamefont {Y.}~\bibnamefont {Matsuda}},\ }\bibfield  {title}
      {\bibinfo {title} {{Half}-integer quantized anomalous thermal {Hall} effect
      in the {Kitaev} material candidate $\alpha$-{RuCl}$_3$},\ }\href
      {https://doi.org/10.1126/science.aay5551} {\bibfield  {journal} {\bibinfo
      {journal} {Science}\ }\textbf {\bibinfo {volume} {373}},\ \bibinfo {pages}
      {568} (\bibinfo {year} {2021})}\BibitemShut {NoStop}%
    \bibitem [{\citenamefont {Bruin}\ \emph {et~al.}(2021)\citenamefont {Bruin},
      \citenamefont {Claus}, \citenamefont {Matsumoto}, \citenamefont {Kurita},
      \citenamefont {Tanaka},\ and\ \citenamefont
      {Takagi}}]{bruin_robustness_2021}%
      \BibitemOpen
      \bibfield  {author} {\bibinfo {author} {\bibfnamefont {J.~A.~N.}\
      \bibnamefont {Bruin}}, \bibinfo {author} {\bibfnamefont {R.~R.}\ \bibnamefont
      {Claus}}, \bibinfo {author} {\bibfnamefont {Y.}~\bibnamefont {Matsumoto}},
      \bibinfo {author} {\bibfnamefont {N.}~\bibnamefont {Kurita}}, \bibinfo
      {author} {\bibfnamefont {H.}~\bibnamefont {Tanaka}},\ and\ \bibinfo {author}
      {\bibfnamefont {H.}~\bibnamefont {Takagi}},\ }\bibfield  {title} {\bibinfo
      {title} {{Robustness} of the thermal {Hall} effect close to half-quantization
      in a field-induced spin liquid state},\ }\href@noop {} {\bibfield  {journal}
      {\bibinfo  {journal} {arXiv:2104.12184}\ } (\bibinfo {year}
      {2021})}\BibitemShut {NoStop}%
    \bibitem [{\citenamefont {Grissonnanche}\ \emph {et~al.}(2019)\citenamefont
      {Grissonnanche}, \citenamefont {Legros}, \citenamefont {Badoux},
      \citenamefont {Lefran\c{c}ois}, \citenamefont {Zatko}, \citenamefont
      {Lizaire}, \citenamefont {Lalibert\'e}, \citenamefont {Gourgout},
      \citenamefont {Zhou}, \citenamefont {Pyon}, \citenamefont {Takayama},
      \citenamefont {Takagi}, \citenamefont {Ono}, \citenamefont {Doiron-Leyraud},\
      and\ \citenamefont {Taillefer}}]{grissonnanche_giant_2019}%
      \BibitemOpen
      \bibfield  {author} {\bibinfo {author} {\bibfnamefont {G.}~\bibnamefont
      {Grissonnanche}}, \bibinfo {author} {\bibfnamefont {A.}~\bibnamefont
      {Legros}}, \bibinfo {author} {\bibfnamefont {S.}~\bibnamefont {Badoux}},
      \bibinfo {author} {\bibfnamefont {E.}~\bibnamefont {Lefran\c{c}ois}},
      \bibinfo {author} {\bibfnamefont {V.}~\bibnamefont {Zatko}}, \bibinfo
      {author} {\bibfnamefont {M.}~\bibnamefont {Lizaire}}, \bibinfo {author}
      {\bibfnamefont {F.}~\bibnamefont {Lalibert\'e}}, \bibinfo {author}
      {\bibfnamefont {A.}~\bibnamefont {Gourgout}}, \bibinfo {author}
      {\bibfnamefont {J.-S.}\ \bibnamefont {Zhou}}, \bibinfo {author}
      {\bibfnamefont {S.}~\bibnamefont {Pyon}}, \bibinfo {author} {\bibfnamefont
      {T.}~\bibnamefont {Takayama}}, \bibinfo {author} {\bibfnamefont
      {H.}~\bibnamefont {Takagi}}, \bibinfo {author} {\bibfnamefont
      {S.}~\bibnamefont {Ono}}, \bibinfo {author} {\bibfnamefont {N.}~\bibnamefont
      {Doiron-Leyraud}},\ and\ \bibinfo {author} {\bibfnamefont {L.}~\bibnamefont
      {Taillefer}},\ }\bibfield  {title} {\bibinfo {title} {{Giant} thermal {Hall}
      conductivity in the pseudogap phase of cuprate superconductors},\ }\href
      {https://doi.org/10.1038/s41586-019-1375-0} {\bibfield  {journal} {\bibinfo
      {journal} {Nature}\ }\textbf {\bibinfo {volume} {571}},\ \bibinfo {pages}
      {376} (\bibinfo {year} {2019})}\BibitemShut {NoStop}%
    \bibitem [{\citenamefont {Hirokane}\ \emph {et~al.}(2019)\citenamefont
      {Hirokane}, \citenamefont {Nii}, \citenamefont {Tomioka},\ and\ \citenamefont
      {Onose}}]{hirokane_phononic_2019}%
      \BibitemOpen
      \bibfield  {author} {\bibinfo {author} {\bibfnamefont {Y.}~\bibnamefont
      {Hirokane}}, \bibinfo {author} {\bibfnamefont {Y.}~\bibnamefont {Nii}},
      \bibinfo {author} {\bibfnamefont {Y.}~\bibnamefont {Tomioka}},\ and\ \bibinfo
      {author} {\bibfnamefont {Y.}~\bibnamefont {Onose}},\ }\bibfield  {title}
      {\bibinfo {title} {{Phononic} thermal {Hall} effect in diluted terbium
      oxides},\ }\href {https://doi.org/10.1103/PhysRevB.99.134419} {\bibfield
      {journal} {\bibinfo  {journal} {Physical Review B}\ }\textbf {\bibinfo
      {volume} {99}},\ \bibinfo {pages} {134419} (\bibinfo {year}
      {2019})}\BibitemShut {NoStop}%
    \bibitem [{\citenamefont {Li}\ \emph {et~al.}(2020)\citenamefont {Li},
      \citenamefont {Fauqu\'e}, \citenamefont {Zhu},\ and\ \citenamefont
      {Behnia}}]{li_phonon_2020}%
      \BibitemOpen
      \bibfield  {author} {\bibinfo {author} {\bibfnamefont {X.}~\bibnamefont
      {Li}}, \bibinfo {author} {\bibfnamefont {B.}~\bibnamefont {Fauqu\'e}},
      \bibinfo {author} {\bibfnamefont {Z.}~\bibnamefont {Zhu}},\ and\ \bibinfo
      {author} {\bibfnamefont {K.}~\bibnamefont {Behnia}},\ }\bibfield  {title}
      {\bibinfo {title} {{Phonon} {Thermal} {Hall} {Effect} in {Strontium}
      {Titanate}},\ }\href {https://doi.org/10.1103/PhysRevLett.124.105901}
      {\bibfield  {journal} {\bibinfo  {journal} {Physical Review Letters}\
      }\textbf {\bibinfo {volume} {124}},\ \bibinfo {pages} {105901} (\bibinfo
      {year} {2020})}\BibitemShut {NoStop}%
    \bibitem [{\citenamefont {Grissonnanche}\ \emph {et~al.}(2020)\citenamefont
      {Grissonnanche}, \citenamefont {Th\'eriault}, \citenamefont {Gourgout},
      \citenamefont {Boulanger}, \citenamefont {Lefran\c{c}ois}, \citenamefont
      {Ataei}, \citenamefont {Lalibert\'e}, \citenamefont {Dion}, \citenamefont
      {Zhou}, \citenamefont {Pyon}, \citenamefont {Takayama}, \citenamefont
      {Takagi}, \citenamefont {Doiron-Leyraud},\ and\ \citenamefont
      {Taillefer}}]{grissonnanche_chiral_2020}%
      \BibitemOpen
      \bibfield  {author} {\bibinfo {author} {\bibfnamefont {G.}~\bibnamefont
      {Grissonnanche}}, \bibinfo {author} {\bibfnamefont {S.}~\bibnamefont
      {Th\'eriault}}, \bibinfo {author} {\bibfnamefont {A.}~\bibnamefont
      {Gourgout}}, \bibinfo {author} {\bibfnamefont {M.-E.}\ \bibnamefont
      {Boulanger}}, \bibinfo {author} {\bibfnamefont {E.}~\bibnamefont
      {Lefran\c{c}ois}}, \bibinfo {author} {\bibfnamefont {A.}~\bibnamefont
      {Ataei}}, \bibinfo {author} {\bibfnamefont {F.}~\bibnamefont {Lalibert\'e}},
      \bibinfo {author} {\bibfnamefont {M.}~\bibnamefont {Dion}}, \bibinfo {author}
      {\bibfnamefont {J.-S.}\ \bibnamefont {Zhou}}, \bibinfo {author}
      {\bibfnamefont {S.}~\bibnamefont {Pyon}}, \bibinfo {author} {\bibfnamefont
      {T.}~\bibnamefont {Takayama}}, \bibinfo {author} {\bibfnamefont
      {H.}~\bibnamefont {Takagi}}, \bibinfo {author} {\bibfnamefont
      {N.}~\bibnamefont {Doiron-Leyraud}},\ and\ \bibinfo {author} {\bibfnamefont
      {L.}~\bibnamefont {Taillefer}},\ }\bibfield  {title} {\bibinfo {title}
      {{Chiral} phonons in the pseudogap phase of cuprates},\ }\href
      {https://doi.org/10.1038/s41567-020-0965-y} {\bibfield  {journal} {\bibinfo
      {journal} {Nature Physics}\ }\textbf {\bibinfo {volume} {16}},\ \bibinfo
      {pages} {1108–1111} (\bibinfo {year} {2020})}\BibitemShut {NoStop}%
    \bibitem [{\citenamefont {Boulanger}\ \emph {et~al.}(2020)\citenamefont
      {Boulanger}, \citenamefont {Grissonnanche}, \citenamefont {Badoux},
      \citenamefont {Allaire}, \citenamefont {Lefran\c{c}ois}, \citenamefont
      {Legros}, \citenamefont {Gourgout}, \citenamefont {Dion}, \citenamefont
      {Wang}, \citenamefont {Chen}, \citenamefont {Liang}, \citenamefont {Hardy},
      \citenamefont {Bonn},\ and\ \citenamefont
      {Taillefer}}]{boulanger_thermal_2020}%
      \BibitemOpen
      \bibfield  {author} {\bibinfo {author} {\bibfnamefont {M.-E.}\ \bibnamefont
      {Boulanger}}, \bibinfo {author} {\bibfnamefont {G.}~\bibnamefont
      {Grissonnanche}}, \bibinfo {author} {\bibfnamefont {S.}~\bibnamefont
      {Badoux}}, \bibinfo {author} {\bibfnamefont {A.}~\bibnamefont {Allaire}},
      \bibinfo {author} {\bibfnamefont {E.}~\bibnamefont {Lefran\c{c}ois}},
      \bibinfo {author} {\bibfnamefont {A.}~\bibnamefont {Legros}}, \bibinfo
      {author} {\bibfnamefont {A.}~\bibnamefont {Gourgout}}, \bibinfo {author}
      {\bibfnamefont {M.}~\bibnamefont {Dion}}, \bibinfo {author} {\bibfnamefont
      {C.~H.}\ \bibnamefont {Wang}}, \bibinfo {author} {\bibfnamefont {X.~H.}\
      \bibnamefont {Chen}}, \bibinfo {author} {\bibfnamefont {R.}~\bibnamefont
      {Liang}}, \bibinfo {author} {\bibfnamefont {W.~N.}\ \bibnamefont {Hardy}},
      \bibinfo {author} {\bibfnamefont {D.~A.}\ \bibnamefont {Bonn}},\ and\
      \bibinfo {author} {\bibfnamefont {L.}~\bibnamefont {Taillefer}},\ }\bibfield
      {title} {\bibinfo {title} {{Thermal} {Hall} conductivity in the cuprate
      {Mott} insulators {Nd}$_2${CuO}$_4$ and {Sr}$_2${CuO}$_2${Cl}$_2$},\ }\href
      {https://doi.org/10.1038/s41467-020-18881-z} {\bibfield  {journal} {\bibinfo
      {journal} {Nature Communications}\ }\textbf {\bibinfo {volume} {11}},\
      \bibinfo {pages} {5325} (\bibinfo {year} {2020})}\BibitemShut {NoStop}%
    \bibitem [{\citenamefont {May}\ \emph {et~al.}(2020)\citenamefont {May},
      \citenamefont {Yan},\ and\ \citenamefont {McGuire}}]{may_practical_2020}%
      \BibitemOpen
      \bibfield  {author} {\bibinfo {author} {\bibfnamefont {A.~F.}\ \bibnamefont
      {May}}, \bibinfo {author} {\bibfnamefont {J.}~\bibnamefont {Yan}},\ and\
      \bibinfo {author} {\bibfnamefont {M.~A.}\ \bibnamefont {McGuire}},\
      }\bibfield  {title} {\bibinfo {title} {{A} practical guide for crystal growth
      of van der {Waals} layered materials},\ }\href
      {https://doi.org/10.1063/5.0015971} {\bibfield  {journal} {\bibinfo
      {journal} {Journal of Applied Physics}\ }\textbf {\bibinfo {volume} {128}},\
      \bibinfo {pages} {051101} (\bibinfo {year} {2020})}\BibitemShut {NoStop}%
    \bibitem [{\citenamefont {Sears}\ \emph {et~al.}(2017)\citenamefont {Sears},
      \citenamefont {Zhao}, \citenamefont {Xu}, \citenamefont {Lynn},\ and\
      \citenamefont {Kim}}]{sears_phase_2017}%
      \BibitemOpen
      \bibfield  {author} {\bibinfo {author} {\bibfnamefont {J.~A.}\ \bibnamefont
      {Sears}}, \bibinfo {author} {\bibfnamefont {Y.}~\bibnamefont {Zhao}},
      \bibinfo {author} {\bibfnamefont {Z.}~\bibnamefont {Xu}}, \bibinfo {author}
      {\bibfnamefont {J.~W.}\ \bibnamefont {Lynn}},\ and\ \bibinfo {author}
      {\bibfnamefont {Y.-J.}\ \bibnamefont {Kim}},\ }\bibfield  {title} {\bibinfo
      {title} {{Phase} diagram of $\alpha$-{RuCl}$_3$ in an in-plane magnetic
      field},\ }\href {https://doi.org/10.1103/PhysRevB.95.180411} {\bibfield
      {journal} {\bibinfo  {journal} {Physical Review B}\ }\textbf {\bibinfo
      {volume} {95}},\ \bibinfo {pages} {180411} (\bibinfo {year}
      {2017})}\BibitemShut {NoStop}%
    \bibitem [{\citenamefont {Leahy}\ \emph {et~al.}(2017)\citenamefont {Leahy},
      \citenamefont {Pocs}, \citenamefont {Siegfried}, \citenamefont {Graf},
      \citenamefont {Do}, \citenamefont {Choi}, \citenamefont {Normand},\ and\
      \citenamefont {Lee}}]{leahy_anomalous_2017}%
      \BibitemOpen
      \bibfield  {author} {\bibinfo {author} {\bibfnamefont {I.~A.}\ \bibnamefont
      {Leahy}}, \bibinfo {author} {\bibfnamefont {C.~A.}\ \bibnamefont {Pocs}},
      \bibinfo {author} {\bibfnamefont {P.~E.}\ \bibnamefont {Siegfried}}, \bibinfo
      {author} {\bibfnamefont {D.}~\bibnamefont {Graf}}, \bibinfo {author}
      {\bibfnamefont {S.-H.}\ \bibnamefont {Do}}, \bibinfo {author} {\bibfnamefont
      {K.-Y.}\ \bibnamefont {Choi}}, \bibinfo {author} {\bibfnamefont
      {B.}~\bibnamefont {Normand}},\ and\ \bibinfo {author} {\bibfnamefont
      {M.}~\bibnamefont {Lee}},\ }\bibfield  {title} {\bibinfo {title} {{Anomalous}
      {Thermal} {Conductivity} and {Magnetic} {Torque} {Response} in the
      {Honeycomb} {Magnet} $\alpha$-{RuCl}$_3$},\ }\href
      {https://doi.org/10.1103/PhysRevLett.118.187203} {\bibfield  {journal}
      {\bibinfo  {journal} {Physical Review Letters}\ }\textbf {\bibinfo {volume}
      {118}},\ \bibinfo {pages} {187203} (\bibinfo {year} {2017})}\BibitemShut
      {NoStop}%
    \bibitem [{\citenamefont {Hentrich}\ \emph {et~al.}(2018)\citenamefont
      {Hentrich}, \citenamefont {Wolter}, \citenamefont {Zotos}, \citenamefont
      {Brenig}, \citenamefont {Nowak}, \citenamefont {Isaeva}, \citenamefont
      {Doert}, \citenamefont {Banerjee}, \citenamefont {Lampen-Kelley},
      \citenamefont {Mandrus}, \citenamefont {Nagler}, \citenamefont {Sears},
      \citenamefont {Kim}, \citenamefont {B\"uchner},\ and\ \citenamefont
      {Hess}}]{hentrich_unusual_2018}%
      \BibitemOpen
      \bibfield  {author} {\bibinfo {author} {\bibfnamefont {R.}~\bibnamefont
      {Hentrich}}, \bibinfo {author} {\bibfnamefont {A.~U.}\ \bibnamefont
      {Wolter}}, \bibinfo {author} {\bibfnamefont {X.}~\bibnamefont {Zotos}},
      \bibinfo {author} {\bibfnamefont {W.}~\bibnamefont {Brenig}}, \bibinfo
      {author} {\bibfnamefont {D.}~\bibnamefont {Nowak}}, \bibinfo {author}
      {\bibfnamefont {A.}~\bibnamefont {Isaeva}}, \bibinfo {author} {\bibfnamefont
      {T.}~\bibnamefont {Doert}}, \bibinfo {author} {\bibfnamefont
      {A.}~\bibnamefont {Banerjee}}, \bibinfo {author} {\bibfnamefont
      {P.}~\bibnamefont {Lampen-Kelley}}, \bibinfo {author} {\bibfnamefont {D.~G.}\
      \bibnamefont {Mandrus}}, \bibinfo {author} {\bibfnamefont {S.~E.}\
      \bibnamefont {Nagler}}, \bibinfo {author} {\bibfnamefont {J.}~\bibnamefont
      {Sears}}, \bibinfo {author} {\bibfnamefont {Y.-J.}\ \bibnamefont {Kim}},
      \bibinfo {author} {\bibfnamefont {B.}~\bibnamefont {B\"uchner}},\ and\
      \bibinfo {author} {\bibfnamefont {C.}~\bibnamefont {Hess}},\ }\bibfield
      {title} {\bibinfo {title} {{Unusual} {Phonon} {Heat} {Transport} in
      $\alpha$-{RuCl}$_3$: {Strong} {Spin}-{Phonon} {Scattering} and
      {Field}-{Induced} {Spin} {Gap}},\ }\href
      {https://doi.org/10.1103/PhysRevLett.120.117204} {\bibfield  {journal}
      {\bibinfo  {journal} {Physical Review Letters}\ }\textbf {\bibinfo {volume}
      {120}},\ \bibinfo {pages} {117204} (\bibinfo {year} {2018})}\BibitemShut
      {NoStop}%
    \bibitem [{\citenamefont {Kasahara}\ \emph
      {et~al.}(2018{\natexlab{b}})\citenamefont {Kasahara}, \citenamefont {Sugii},
      \citenamefont {Ohnishi}, \citenamefont {Shimozawa}, \citenamefont
      {Yamashita}, \citenamefont {Kurita}, \citenamefont {Tanaka}, \citenamefont
      {Nasu}, \citenamefont {Motome}, \citenamefont {Shibauchi},\ and\
      \citenamefont {Matsuda}}]{kasahara_unusual_2018}%
      \BibitemOpen
      \bibfield  {author} {\bibinfo {author} {\bibfnamefont {Y.}~\bibnamefont
      {Kasahara}}, \bibinfo {author} {\bibfnamefont {K.}~\bibnamefont {Sugii}},
      \bibinfo {author} {\bibfnamefont {T.}~\bibnamefont {Ohnishi}}, \bibinfo
      {author} {\bibfnamefont {M.}~\bibnamefont {Shimozawa}}, \bibinfo {author}
      {\bibfnamefont {M.}~\bibnamefont {Yamashita}}, \bibinfo {author}
      {\bibfnamefont {N.}~\bibnamefont {Kurita}}, \bibinfo {author} {\bibfnamefont
      {H.}~\bibnamefont {Tanaka}}, \bibinfo {author} {\bibfnamefont
      {J.}~\bibnamefont {Nasu}}, \bibinfo {author} {\bibfnamefont {Y.}~\bibnamefont
      {Motome}}, \bibinfo {author} {\bibfnamefont {T.}~\bibnamefont {Shibauchi}},\
      and\ \bibinfo {author} {\bibfnamefont {Y.}~\bibnamefont {Matsuda}},\
      }\bibfield  {title} {\bibinfo {title} {{Unusual} {Thermal} {Hall} {Effect} in
      a {Kitaev} {Spin} {Liquid} {Candidate} $\alpha$-{RuCl}$_3$},\ }\href
      {https://doi.org/10.1103/PhysRevLett.120.217205} {\bibfield  {journal}
      {\bibinfo  {journal} {Physical Review Letters}\ }\textbf {\bibinfo {volume}
      {120}},\ \bibinfo {pages} {217205} (\bibinfo {year}
      {2018}{\natexlab{b}})}\BibitemShut {NoStop}%
    \bibitem [{\citenamefont {Yu}\ \emph {et~al.}(2018)\citenamefont {Yu},
      \citenamefont {Xu}, \citenamefont {Ran}, \citenamefont {Ni}, \citenamefont
      {Huang}, \citenamefont {Wang}, \citenamefont {Wen},\ and\ \citenamefont
      {Li}}]{yu_ultralow-temperature_2018}%
      \BibitemOpen
      \bibfield  {author} {\bibinfo {author} {\bibfnamefont {Y.}~\bibnamefont
      {Yu}}, \bibinfo {author} {\bibfnamefont {Y.}~\bibnamefont {Xu}}, \bibinfo
      {author} {\bibfnamefont {K.}~\bibnamefont {Ran}}, \bibinfo {author}
      {\bibfnamefont {J.}~\bibnamefont {Ni}}, \bibinfo {author} {\bibfnamefont
      {Y.}~\bibnamefont {Huang}}, \bibinfo {author} {\bibfnamefont
      {J.}~\bibnamefont {Wang}}, \bibinfo {author} {\bibfnamefont {J.}~\bibnamefont
      {Wen}},\ and\ \bibinfo {author} {\bibfnamefont {S.}~\bibnamefont {Li}},\
      }\bibfield  {title} {\bibinfo {title} {{Ultralow}-{Temperature} {Thermal}
      {Conductivity} of the {Kitaev} {Honeycomb} {Magnet} $\alpha$-{RuCl}$_3$
      across the {Field}-{Induced} {Phase} {Transition}},\ }\href
      {https://doi.org/10.1103/PhysRevLett.120.067202} {\bibfield  {journal}
      {\bibinfo  {journal} {Physical Review Letters}\ }\textbf {\bibinfo {volume}
      {120}},\ \bibinfo {pages} {067202} (\bibinfo {year} {2018})}\BibitemShut
      {NoStop}%
    \bibitem [{\citenamefont {Hentrich}\ \emph {et~al.}(2019)\citenamefont
      {Hentrich}, \citenamefont {Roslova}, \citenamefont {Isaeva}, \citenamefont
      {Doert}, \citenamefont {Brenig}, \citenamefont {B\"uchner},\ and\
      \citenamefont {Hess}}]{hentrich_large_2019}%
      \BibitemOpen
      \bibfield  {author} {\bibinfo {author} {\bibfnamefont {R.}~\bibnamefont
      {Hentrich}}, \bibinfo {author} {\bibfnamefont {M.}~\bibnamefont {Roslova}},
      \bibinfo {author} {\bibfnamefont {A.}~\bibnamefont {Isaeva}}, \bibinfo
      {author} {\bibfnamefont {T.}~\bibnamefont {Doert}}, \bibinfo {author}
      {\bibfnamefont {W.}~\bibnamefont {Brenig}}, \bibinfo {author} {\bibfnamefont
      {B.}~\bibnamefont {B\"uchner}},\ and\ \bibinfo {author} {\bibfnamefont
      {C.}~\bibnamefont {Hess}},\ }\bibfield  {title} {\bibinfo {title} {{Large}
      thermal {Hall} effect in $\alpha$-{RuCl}$_3$: {Evidence} for heat transport
      by {Kitaev}-{Heisenberg} paramagnons},\ }\href
      {https://doi.org/10.1103/PhysRevB.99.085136} {\bibfield  {journal} {\bibinfo
      {journal} {Physical Review B}\ }\textbf {\bibinfo {volume} {99}},\ \bibinfo
      {pages} {085136} (\bibinfo {year} {2019})}\BibitemShut {NoStop}%
    \bibitem [{\citenamefont {Hirschberger}\ \emph {et~al.}(2015)\citenamefont
      {Hirschberger}, \citenamefont {Chisnell}, \citenamefont {Lee},\ and\
      \citenamefont {Ong}}]{hirschberger_thermal_2015}%
      \BibitemOpen
      \bibfield  {author} {\bibinfo {author} {\bibfnamefont {M.}~\bibnamefont
      {Hirschberger}}, \bibinfo {author} {\bibfnamefont {R.}~\bibnamefont
      {Chisnell}}, \bibinfo {author} {\bibfnamefont {Y.~S.}\ \bibnamefont {Lee}},\
      and\ \bibinfo {author} {\bibfnamefont {N.}~\bibnamefont {Ong}},\ }\bibfield
      {title} {\bibinfo {title} {{Thermal} {Hall} {Effect} of {Spin} {Excitations}
      in a {Kagome} {Magnet}},\ }\href
      {https://doi.org/10.1103/PhysRevLett.115.106603} {\bibfield  {journal}
      {\bibinfo  {journal} {Physical Review Letters}\ }\textbf {\bibinfo {volume}
      {115}},\ \bibinfo {pages} {106603} (\bibinfo {year} {2015})}\BibitemShut
      {NoStop}%
    \bibitem [{\citenamefont {Strohm}\ \emph {et~al.}(2005)\citenamefont {Strohm},
      \citenamefont {Rikken},\ and\ \citenamefont
      {Wyder}}]{strohm_phenomenological_2005}%
      \BibitemOpen
      \bibfield  {author} {\bibinfo {author} {\bibfnamefont {C.}~\bibnamefont
      {Strohm}}, \bibinfo {author} {\bibfnamefont {G.~L. J.~A.}\ \bibnamefont
      {Rikken}},\ and\ \bibinfo {author} {\bibfnamefont {P.}~\bibnamefont
      {Wyder}},\ }\bibfield  {title} {\bibinfo {title} {{Phenomenological}
      {Evidence} for the {Phonon} {Hall} {Effect}},\ }\href
      {https://doi.org/10.1103/PhysRevLett.95.155901} {\bibfield  {journal}
      {\bibinfo  {journal} {Physical Review Letters}\ }\textbf {\bibinfo {volume}
      {95}},\ \bibinfo {pages} {155901} (\bibinfo {year} {2005})}\BibitemShut
      {NoStop}%
    \bibitem [{\citenamefont {Ideue}\ \emph {et~al.}(2017)\citenamefont {Ideue},
      \citenamefont {Kurumaji}, \citenamefont {Ishiwata},\ and\ \citenamefont
      {Tokura}}]{ideue_giant_2017}%
      \BibitemOpen
      \bibfield  {author} {\bibinfo {author} {\bibfnamefont {T.}~\bibnamefont
      {Ideue}}, \bibinfo {author} {\bibfnamefont {T.}~\bibnamefont {Kurumaji}},
      \bibinfo {author} {\bibfnamefont {S.}~\bibnamefont {Ishiwata}},\ and\
      \bibinfo {author} {\bibfnamefont {Y.}~\bibnamefont {Tokura}},\ }\bibfield
      {title} {\bibinfo {title} {{Giant} thermal {Hall} effect in multiferroics},\
      }\href {https://doi.org/10.1038/nmat4905} {\bibfield  {journal} {\bibinfo
      {journal} {Nature Materials}\ }\textbf {\bibinfo {volume} {16}},\ \bibinfo
      {pages} {797} (\bibinfo {year} {2017})}\BibitemShut {NoStop}%
    \bibitem [{\citenamefont {Chen}\ \emph {et~al.}(2021)\citenamefont {Chen},
      \citenamefont {Boulanger}, \citenamefont {Wang}, \citenamefont {Tafti},\ and\
      \citenamefont {Taillefer}}]{chen_large_2021}%
      \BibitemOpen
      \bibfield  {author} {\bibinfo {author} {\bibfnamefont {L.}~\bibnamefont
      {Chen}}, \bibinfo {author} {\bibfnamefont {M.-E.}\ \bibnamefont {Boulanger}},
      \bibinfo {author} {\bibfnamefont {Z.-C.}\ \bibnamefont {Wang}}, \bibinfo
      {author} {\bibfnamefont {F.}~\bibnamefont {Tafti}},\ and\ \bibinfo {author}
      {\bibfnamefont {L.}~\bibnamefont {Taillefer}},\ }\bibfield  {title} {\bibinfo
      {title} {{Large} {Phonon} {Thermal} {Hall} {Conductivity} in a {Simple}
      {Antiferromagnetic} {Insulator}},\ }\href@noop {} {\bibfield  {journal}
      {\bibinfo  {journal} {arXiv:2110.13277}\ } (\bibinfo {year}
      {2021})}\BibitemShut {NoStop}%
    \bibitem [{\citenamefont {Yamashita}\ \emph {et~al.}(2020)\citenamefont
      {Yamashita}, \citenamefont {Gouchi}, \citenamefont {Uwatoko}, \citenamefont
      {Kurita},\ and\ \citenamefont {Tanaka}}]{yamashita_sample_2020}%
      \BibitemOpen
      \bibfield  {author} {\bibinfo {author} {\bibfnamefont {M.}~\bibnamefont
      {Yamashita}}, \bibinfo {author} {\bibfnamefont {J.}~\bibnamefont {Gouchi}},
      \bibinfo {author} {\bibfnamefont {Y.}~\bibnamefont {Uwatoko}}, \bibinfo
      {author} {\bibfnamefont {N.}~\bibnamefont {Kurita}},\ and\ \bibinfo {author}
      {\bibfnamefont {H.}~\bibnamefont {Tanaka}},\ }\bibfield  {title} {\bibinfo
      {title} {{Sample} dependence of half-integer quantized thermal {Hall} effect
      in the {Kitaev} spin-liquid candidate $\alpha$-{RuCl}$_3$},\ }\href
      {https://doi.org/10.1103/PhysRevB.102.220404} {\bibfield  {journal} {\bibinfo
       {journal} {Physical Review B}\ }\textbf {\bibinfo {volume} {102}},\ \bibinfo
      {pages} {220404} (\bibinfo {year} {2020})}\BibitemShut {NoStop}%
    \end{thebibliography}
    
%

\end{document}